\documentclass[aps,prb,twocolumn,showpacs]{revtex4}%
\usepackage{amsmath}
\usepackage{graphicx}
\usepackage{epsfig}
\usepackage{amsfonts}
\usepackage{amssymb}%
\providecommand{\U}[1]{\protect\rule{.1in}{.1in}}
\begin{document}
\title{ About the role of 2D screening in High Temperature Superconductivity}
\author{Yosdanis Vazquez-Ponce }
\thanks{}
\email{ yvponce@gmail.com}
 \affiliation{Group of Theoretical Physics, Instituto de
Cibern\'etica, Matem\'atica y F\'{\i}sica, Calle E, No. 309,
Vedado, La Habana, Cuba}
\author{ David Oliva Aguero }
\thanks{}
\email{david@cidet.icmf.inf.cu} \affiliation{Group of Theoretical
Physics, Instituto de Cibern\'etica, Matem\'atica y F\'{\i}sica,
Calle E, No. 309, Vedado, La Habana, Cuba}

\author{Alejandro Cabo Montes de Oca }
\thanks{}
\email{cabo@cidet.icmf.inf.cu}
 \altaffiliation[Permanent address:] {\ Group of Theoretical
Physics, Instituto de Cibern\'etica, Matem\'atica y F\'{\i}sica,
Calle E, No. 309, Vedado, \ La Habana, Cuba.}
\affiliation{International Centre for Theoretical Physics, Strada
Costiera 11-34014, Miramare, Trieste, Italy.}

\begin{abstract}
\noindent The 2D screening is investigated in a simple single band
square tight-binding model which qualitatively resembles the known
electronic structure in high temperature superconductors. The
Coulomb kernel for the two particle Bethe-Salpeter equation in the
single loop (RPA) approximation for the polarization \ can be
evaluated in a strong tight binding limit. The results indicate an
intense screening of the Coulomb repulsion between the particles,
which becomes stronger and anisotropic when the Fermi level
approach half filling (or equivalently, when the Fermi surface
approach the Van Hove singularities) and rapidly decreases away
it. The effect is also more pronounced for quasi-momenta regions
near the corners of the Brillouin cell, which correspond to dual
spatial distances of the order few unit cells. Therefore, a
possible mechanism is identified which could explain the existence
of extremely small Cooper pairs in these materials, as bounded
anisotropic composites joined by residual super-exchange or phonon
interactions.
\end{abstract}


\maketitle

\section{Introduction}

The small dimension of the Cooper pairs in the High Temperature (HTc)
superconductors remains being an unclear point in the physics of these
materials \cite{dagotto,scalapino,johnston}. In our view the main point to
understand is the way in which the strong Coulomb repulsion which should be
present between the two particles at distances of the order $15$ $\mathring
{A}$ (the estimated size of the pairs), should be somehow compensated for
allowing the particles to be bounded. The cancellation in the old
superconducting materials is carried out by the normal metallic screening at
the Cooper pair sizes in these materials which are of the order of hundreds of
$\mathring{A}$. It allows the weak phonon attraction to realize the binding.

Considering this situation one is led to the idea that there should exist
special characteristics of the HTc materials allowing the mentioned
compensation of the Coulomb repulsion. One of the most salient features of the
materials showing high Tc superconductivity is the presence of lattice planes
formed by Cu and O atoms in a nearly square arrangement \cite{mattheiss}.
Therefore, it can be naturally suspected that the 2D character of the electron
dynamics of those planes could be closely connected with the screening of the
Coulomb repulsion at short distances. Following this idea, in the present work
we investigate the screening of the Coulomb potential in a simple tight
binding model chosen to qualitatively resemble the valence electron properties
of the high Tc materials. In order to allow the arrival to analytical results,
the model is constructed as simple as possible and purely two dimensional.

The main aim will be to obtain an estimation of the dielectric
function and the kernel of the Bethe-Salpeter bound state equation
associated to the Coulomb interaction. The model  will be a simple
tight binding one designed, roughly speaking, to match the
dispersion relation of the half filled band crossed by the Fermi
energy in the high Tc materials. The energetic width of this band
is extracted from the work [\onlinecite{mattheiss}]. \ \ \ Some
additional simplifying assumptions also help to reduce the
mathematical complications in the estimation of the dielectric
function and the kernel. However, we think that they do not
sacrifice the qualitative physical essence of the discussion. It
should be stressed that the justification for starting from this
simple model, comes from the need of reducing the technical
complications created by the lack of full translation invariance
of the physical problem. This fact cancels the simple dependence
of the dielectric response quantities of homogeneous systems on
the difference of the spatial arguments. This circumstance led us
to search for analytical results in the non-interacting part of
the electronic Hamiltonian allowing to better tackling the
complications in evaluating the dielectric response. The
calculations are done  following the method developed by Hanke and
Sham for the calculation of the dielectric response properties of
crystalline systems \cite{hanke,sham,dolgov}.

In our view the obtained results for the dielectric function and
the interaction kernel \ have interesting physical implications.
They indicate that, whenever other higher contributions do not
balance out the effect of the here presented ones, the amount of
screening produced by the planar electrons  can be intense when
the system is near half filling. Thus, the possibility is
suggested of the occurrence of a strong screening of the Coulomb
repulsion in the real materials for small doping. Being close half
filling, the evaluated dielectric functions can reduce the value
of few $eV $ of the bare Coulomb potential at distances of few
periods of the spatial lattice, to values of the order of 0.1
$eV$. Therefore, a possibility is indicated for other weak binding
mechanisms to furnish the necessary attractive forces for the pair
creation. Current ideas consider that a combination of diverse
forces (super-exchange, electron-phonon, excitonic, polaronic,
etc.) could play a role in the formation of the Cooper pairs.
However, experiments have still not allowed to fix the \ nature of
the acting mechanism, or at least, there is no consensus about the
relevance of any particular one \cite{scalapino,dagotto}. It also
follows that the screening effects rapidly decrease when the
density of holes created on the half filled ground state
increases. This property in competition with the natural need for
a non vanishing hole density for the superconductor condensate to
exist, could furnish an explanation for the first growing and
after decreasing value of the critical temperature as a function
of the hole density.

Up to our knowledge, the idea that the Coulomb potential in a 2D
crystal system can contribute to produce, not only screening but
even binding between pairs, was firstly advanced by D. Mattis
\cite{mattis}. Also various authors have been also considering the
effects of screening in superconductivity along various directions
of thinking \cite{shuttler,vander,cape, belya,shima,koch}.

A special point to be noticed is the fact that the here considered
RPA approximation, in general grounds, is more \ suitable in
situations in which the kinetic energy is greater than the
interaction one \cite{pines}. \ This means that the limit of
validity of the here evaluated quantities should be closely
examined, since the interactions are \ important in the high Tc
materials. However, it can be stressed that the applicability of
the RPA in \ Hubbard like models in the metallic region up to near
the metal-insulator transition has been argued \cite{koch}. This
question is expected to be more closely addressed elsewhere. It
should be also underlined that the results presented here support
the also strong screening effects, and even over screening, of the
on-site Coulomb coefficients in single and two bands Hubbard
models reported in [\onlinecite{shuttler}]. The study of the
possible connections between these results with the ones presented
in this work seems to be worth considering in further studies.

The particular motivation for realization of the present work was
created by the already mentioned observations about the
characteristics of the HTc superconductive systems: a) \ The
remarkable smallness of the Cooper pairs (of the order of $15  \
\mathring{A}$). b) The foreseeable need of a mechanism being
effective in screening the Coulomb repulsion, which is of the
order of few $eV$ at distances of the size of the Cooper pairs.
The existence of such a mechanism then could allow the short
distance attraction of other participating forces, like the
super-exchange interactions by example,  to bind the pairs
\cite{Zhang_Rice}.

The structure of the work will be as follows: \ In Section 2, the
tight-binding model for the $ab$ ceramic $CuO$ planes will be
defined. Simple Gaussian orbitals are introduced from the start,
allowing (following Wannier \cite{wannier}) to find explicit
expressions for the Bloch functions. Then, the tight binding
dispersion relation $\epsilon(k)$ for the model is matched to the
one shown by the existing half filled band in the $HTSC$ materials
\cite{mattheiss}. In Section 3 \ the fermion free propagator of
the model is employed to obtain formulae for the proper
polarization and its Fourier transform in the ladder (RPA)
approximation. Further, in Section 4, employing the results for
the proper polarization, a formula is also derived for the
screened Coulomb potential. All these expressions show the
complicate dependence on two arguments which follows from the lack
of translation invariance of the system. Then, Section 5 starts
the application of the tight binding approximation to simplify the
formula for the screened Coulomb potential. The final Section 6 is
devoted to the calculation of close expressions for the dielectric
function and the Coulomb screened kernel of the bound state
Bethe-Salpeter equation. These quantities arise as functions of
only one argument: the conserved transmitted reduced quasi
momentum in the first Brillouin zone. This is the main technical
result of work, since the breaking of the translational invariance
of the problem \ obstacle the finding of such formulae in general.
The simplifications were a direct consequence of the tight binding
approximation and allowed to plot these quantities in the first
Brillouin zone for reasonable values of the parameter defining the
overlapping in the tight-binding model. The results indicate the
possibility for the existence of a strong screening of the Coulomb
interaction for conditions resembling the ones present in the real
materials and they are commented in this ending section.

\section{The single band tight-binding model}

This section introduces the simple model for the \ $CuO$ planes in
the superconducting ceramics. \ As the superconductive properties
along the $CuO$ \ layers are significantly stronger that in the
perpendicular $c$ axis, the movement will be considered as \ two
dimensional. Then, it will be supposed that the electrons move in
a square 2D lattice of \ unit cell size $p=3.82$ $\ \mathring{A}.$
This value of the unit cell parameter of the $CuO$ \ lattice was
borough from the work \cite{alecu}. In order to simplify the
screening calculations in the considered \ non translational
invariant problem, we will also adopt a picture motivated in the
one band $tj$-model, in which simple Gaussian orbitals are
centered only on the $Cu$ sites. It is expected that the results
for the screening effects on the Coulomb potential \ that are
obtained in the work could \ be good indicators of what occurs in
the real systems.

\begin{figure}[h]
\begin{center}
\epsfig{figure=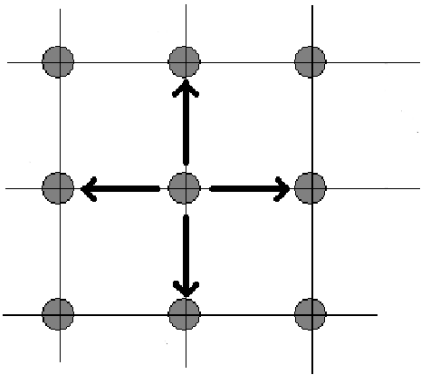,width=6cm}
\end{center}
\caption{ Picture illustrating the square lattice in which the Gaussian
orbitals are centered. The unit cell size is $p=3.82 \mathring{A}$ and the
wave functions are defined on the plane. This simplification allows to obtain
analytic results for some of the screening response quantities. The mean
potential created by the ions and atomic cores is assumed to furnish the
experimentally observed band width of the half filled band crossing the fermi
level. This is another simplifying assumption done in the work. }%
\label{lattice}%
\end{figure}

The Gaussian orbitals staying on the points of the square lattice
(See the figure \ref{lattice}) will have the form
\begin{equation}
\varphi(\mathbf{x})=\frac{1}{\sqrt{N_{at}}} \ e^{-\frac{\left\vert
\mathbf{x}\right\vert ^{2}}{2a^{2}}}\text{,} \ \ N_{at}=\pi a^{2}
,\label{orb_gaus}%
\end{equation}
where $r=\left\vert \mathbf{x}\right\vert $ is the distance to the
origin which will be taken as a fixed point of the lattice and \
$a$ is a positive constant. From now on,  all the vectors \ will
be represented in boldface characters. The function
$\varphi(\mathbf{x})$ is normalized in the whole plane of the
crystal. \ The value of $a$ \  measures the width of the spatial
region where the orbital have appreciable values.

Let us now consider the determination of the analytic form of the
orthonormalized \ set of Bloch wave functions
\cite{wannier,kittel}, \ generated by the orbitals
(\ref{orb_gaus}) when displaced on to all the points of \ the
considered squared lattice. The valence electrons \ of this band
will be assumed to be loosely bounded to the atoms laying on the
lattice points. Then, they can \ propagate along the crystal
feeling a periodic potential created by the ions and the rest of
the electrons being more bounded to the atoms. These more
localized electrons will be assumed to pertain to the filled
energy bands of the whole crystal. \ Then, we will search for the
Bloch functions \ associated to the valence electrons in the form
\
\begin{equation}
\Psi_{\mathbf{k}}(\mathbf{x})=\frac{1}{\sqrt{N_{\mathbf{k}}}}\sum_{\mathbf{R}%
}e^{i\mathbf{k}.\mathbf{R}}\varphi(\mathbf{x}-\mathbf{R})\text{ },
\label{Func_Bloch_planteam}%
\end{equation}
where $\mathbf{R}$ is the position vector of a general point of the square
lattice. As usual, it follows that the functions (\ref{Func_Bloch_planteam})
satisfy the Bloch conditions
\[
\Psi_{\mathbf{k}}(\mathbf{x}+\mathbf{R^{\prime}})=e^{i\mathbf{k}%
.\mathbf{R^{\prime}}}\Psi_{\mathbf{k}}(\mathbf{x})\text{.}%
\]

The proposed Bloch orbitals \ will approximately describe the
valence electrons in the crystal receiving the action of \ the
periodic potential of the atoms and core electrons. \ As stated
before, the rest of the electrons are more deeply bounded to the
atoms \ and their effect will be assumed to produce  produce a
kind of \ mean field influence on the valence electrons.

The just defined functions become orthogonal among them due to the very same
construction, since they are eigenfunctions of the translations in the lattice
with different eigenvalues. It rests only to define the normalization factor
$N_{\mathbf{k}}$ in (\ref{Func_Bloch_planteam}) by imposing the additional
validity of the orthonormality condition
\begin{equation}
\int d^{2}\mathbf{x}\Psi_{\mathbf{k^{\prime}}}^{\ast}(\mathbf{x}%
)\Psi_{\mathbf{k}}(\mathbf{x})=\delta_{\mathbf{k},\mathbf{k^{\prime}}}\text{,}
\label{Orto_Nor_Func_Bloch}%
\end{equation}
where $\delta_{\mathbf{k},\mathbf{k^{\prime}}}$ is the Kronecker
delta in the \ indices $\mathbf{k}$ of the states. \ \ For this
purpose let us substitute the sums defining the Bloch functions in
(\ref{Orto_Nor_Func_Bloch}) and use
the property%

\begin{equation}
\sum_{\mathbf{R}}e^{i(\mathbf{k}-\mathbf{k^{\prime}}).\mathbf{R}}=N\text{
}\delta_{\mathbf{k},\mathbf{k^{\prime}}}\text{, } \label{Rel_Delta_Karman}%
\end{equation}
where $N$ is the total number of points of the lattice and $\mathbf{k}%
=2\pi(\frac{1}{L_{1}}n_{1},\frac{1}{L_{2}}n_{2})$ for which \ $L_{1}$, $L_{2}$
are the spatial periods of the crystal in the two principal orthogonal
directions, defined by the Born-Von Karman boundary conditions. After that,
the normalization constant follows in the form%

\begin{align}
N_{\mathbf{k}}  &  =N\sum_{\mathbf{R}}e^{-i\mathbf{k}.\mathbf{R}}\int
d^{2}\mathbf{x}\varphi(\mathbf{x})\varphi(\mathbf{x}-\mathbf{R}),\text{
}\nonumber\\
&  =N\sum_{\mathbf{R}}e^{-i\mathbf{k}.\mathbf{R}}\int
d^{2}x\varphi
(\mathbf{x})\widehat{T}_{-\mathbf{R}}\varphi(\mathbf{x})\text{ },
\label{Nk_a_medias}%
\end{align}
where $\widehat{T}_{\mathbf{R}}$ \ is the translation operator shifting the
arguments of the functions in the vector $\mathbf{R}$ \ as $\ \widehat
{T}_{\mathbf{R}}\varphi(\mathbf{x})=\varphi(\mathbf{x}+\mathbf{R})$.

The Bloch orbitals can be explicitly evaluated in terms of the Elliptic Theta
functions. It can be noticed from the fact that their definition is given by a
sum of \ Gaussian functions. \ Then, after substituting (\ref{orb_gaus}) in
(\ref{Func_Bloch_planteam}), it follows
\begin{widetext}
\begin{align}
\Psi_{\mathbf{k}}(\mathbf{x})  &  =\frac{1}{\sqrt{N_{\mathbf{k}}}}%
\sum_{\mathbf{R}}e^{i\mathbf{k}.\mathbf{R}}\frac{1}{\sqrt{N_{at}}}%
e^{-\frac{(\mathbf{x}-\mathbf{R})^{2}}{2a^{2}\text{ \ }}}\text{ },\nonumber\\
&  =\frac{1}{\sqrt{N_{\mathbf{k}}N_{at}}}\sum_{n_{1}}e^{-\frac{(n_{1}%
a_{1}-x_{1})^{2}}{2a^{2}}+ik_{1}n_{1}a_{1}}\sum_{n_{2}}e^{-\frac{(n_{2}%
a_{2}-x_{2})^{2}}{2a^{2}}+ik_{2}n_{2}a_{2}}\text{ },\nonumber\\
&  =\frac{1}{\sqrt{N_{\mathbf{k}}N_{at}}}e^{-\frac{\mathbf{x}^{2}}{2a^{2}}%
}\theta_{3}(\frac{a_{1}x_{1}}{2a^{2}i}+\frac{k_{1}a_{1}}{2},q_{1})\theta
_{3}(\frac{a_{2}x_{2}}{2a^{2}i}+\frac{k_{2}a_{2}}{2},q_{2}),
\label{Func_Bloch}%
\end{align}
\end{widetext} where it was defined \
$\mathbf{R}=a_{1}\mathbf{n_{1}}+a_{2}\mathbf{n_{2}}$ and
$\mathbf{x}$ = $x_{1}\mathbf{n_{1}+}$ $x_{2}\mathbf{n_{2}}$\ in
which  $\mathbf{n_{1}}$, $\mathbf{n_{2}}$ are unit vectors along
the directions of the crystal axes. \ The precise definition of
the Elliptic function employed
was%
\begin{equation}
\theta_{3}(u,q)=\sum_{n=-\infty}^{+\infty}q^{n^{2}}e^{2nui}\text{ }.
\label{Func_Eliptica}%
\end{equation}

In an analogous way, the normalization factor can be calculated in the form%
\begin{equation}
N_{\mathbf{k}}=N\theta_{3}(-\frac{k_{1}a_{1}}{2},q_{N_{1}})\theta_{3}%
(-\frac{k_{2}a_{2}}{2},q_{N_{2}})\text{ }. \label{Nk_final}%
\end{equation}

The obtained Bloch orbitals satisfy the property
\begin{equation}
\Psi_{-\mathbf{k}}(\mathbf{x})=\Psi_{\mathbf{k}}^{\ast}(\mathbf{x})\text{ },
\label{Prop_Func_Onda}%
\end{equation}
which directly follows from substituting $\mathbf{k}$ \ by $-\mathbf{k}$ in
(\ref{Func_Bloch}) after taken into account the Elliptic function property:
$\theta_{3}^{\ast}(u,q)=\theta_{3}(-u,q)$ .

Let us impose now the tight binding approximation corresponding to
the physical case in which the overlapping of the valence electron
orbitals is small. \ Therefore, it will be considered that the
wave functions are closely localized \ around the points of the
lattice. Thus, the following approximation for the integral in
(\ref{Nk_a_medias}) can be written
\begin{equation}
\int d^{2}\mathbf{x}\varphi(\mathbf{x})\widehat{T}_{\mathbf{R}}\varphi
(\mathbf{x})\backsimeq\delta_{\mathbf{R},0}\int d^{2}\mathbf{x}\varphi
^{2}(\mathbf{x})=\delta_{\mathbf{R},0}\text{ }. \label{Muy_localizadas}%
\end{equation}

But, substituting (\ref{Muy_localizadas}) in (\ref{Nk_a_medias}) gives
\begin{equation}
N_{\mathbf{k}}=N\sum_{\mathbf{R}}e^{-i\mathbf{k}.\mathbf{R}}\delta
_{-\mathbf{R},0}=N\text{ }. \label{Nk=N}%
\end{equation}
That is, according to (\ref{Muy_localizadas}), it follows \ $N_{\mathbf{k}%
}\backsimeq N$ and the Bloch functions take the simpler form:
\[
\Psi_{\mathbf{k}}(\mathbf{x})=\frac{1}{\sqrt{N}}\sum_{\mathbf{R}%
}e^{i\mathbf{k}.\mathbf{R}}\varphi(\mathbf{x}-\mathbf{R})\text{ }.
\]

Let us now consider the spectrum of the model, assuming that the
valence electron Hamiltonian $\widehat{H}_{e}$ will be the most
relevant one for the description of the $HTc$ superconductor
properties. In other words, \ the action of the ions and the core
electrons will be only considered  as sources of mean external
fields \ acting on the valence electrons. Therefore according to
the Hartree Fock approximation each valence electron will be
moving in the periodic potential \ created by the interaction with
the ions and the core atomic electrons. As the result of these
assumptions, we will remain with a system having as its free
Hamiltonian, the sum of the individual one particle Hamiltonians
of the valence electrons
\begin{equation}
\widehat{H}_{e}=\sum_{i}\widehat{T}_{0i}+\sum_{i}\widehat{U}_{0i}=\sum
_{i}\widehat{H}_{0i}\text{ },
\end{equation}%
\[
\widehat{H}_{0i}=\widehat{T}_{0i}+\widehat{U}_{0i}\text{ },
\]
where $\widehat{H}_{0i}$ is the Hamiltonian of \ a valence electron $i$. \ In
addition, after also considering the tight binding approximation, it will be
supposed that in the neighborhood of each point of the lattice, the free
Hamiltonian of an electron can be well approximated by an atomic Hamiltonian
$\widehat{H}_{at}$ centered in the considered point.

Therefore, under the above specifications the tight-binding
procedure leads to the following  dispersion relation for the
valence electrons in the model
under consideration\cite{mermin}%
\begin{equation}
\varepsilon(\mathbf{k})=\epsilon_{F}-2\gamma\left[  \cos(pk_{x})+\cos
(pk_{y})\right]  \text{ }, \label{Expresion_Espectro_Energia}%
\end{equation}
where $\epsilon_{F}$ is the Fermi energy of the system and $\gamma$ defines
the energetic width of the band as $8 \gamma$.

Thus, the electrons will lay in states described by the wave
functions $\Psi_{\mathbf{k}}(\mathbf{x})$ \ corresponding to
energies given by (\ref{Expresion_Espectro_Energia}) where
$\mathbf{k}=(k_{x},k_{y}).$ Here the wave vectors $\mathbf{k}$
will chosen in the first Brillouin zone by convention,  since
$\mathbf{k}^{\prime}\mathbf{=k+Q}$ correspond to physically
equivalent states. \ \ With the help of
(\ref{Expresion_Espectro_Energia}), it can be checked that
$\varepsilon(\mathbf{k+Q})=\varepsilon(\mathbf{k})$ as the
equivalence implies.

  For the definition of our qualitative model, we simply matched the
square lattice parameter to the experimental value. Also, it was
directly assumed that the one particle potentials created by the
ions and the atomic kernels are the necessary ones for \ fixing
the width of the band in (\ref{Expresion_Espectro_Energia}) to the
estimated value. \ The lattice parameter was the one corresponding
to the $ab$ planes in the ceramic material YBaCuO. \, That is,
$p=a=b=3.82$ \AA \ \ as reported in \cite{alecu}.\ The energy
width of the half filled electronic band in the material, as
defined by the energy difference $8 \gamma$ between the top and
the bottom of the band, was estimated here from the data in \ Ref
$\cite{mattheiss}$ to be $3.7$ $eV$. \ \ \ The relation
(\ref{Expresion_Espectro_Energia}) also well reproduce the form of
the band given in \cite{mattheiss}.

Another assumption chosen  for the sake of simplification is that
the dispersion relation of the half filled band \ is solely
determined by the ionic \ plus atomic kernel potential. It
considers that the interaction between the valence electron does
not strongly contribute to the band width, which is not
necessarily valid \ \ Therefore, we expect in the further
extension of the present work to perform a self consistent
derivation of the band width starting from a more basic
interacting Hamiltonian. \ \

Now, let us precise the form of the free propagator for the
valence electrons to be employed in the further evaluations.
According to the above discussion,
the electron states will have the form%
\[
\Psi_{\mathbf{k}\lambda}(\mathbf{x})=\Psi_{\mathbf{k}}(\mathbf{x}%
)\Phi_{\lambda}\text{ },
\]
where $\Psi_{\mathbf{k}}(\mathbf{x})$ are the Bloch wave functions and
\ $\Phi_{\uparrow}=[%
\genfrac{}{}{0pt}{}{1}{0}%
]$ \ \ y $\Phi_{\downarrow}=[%
\genfrac{}{}{0pt}{}{0}{1}%
]$ are the spinor states. In addition, as defined before, $\mathbf{k}%
=2\pi\left(  \frac{n_{1}}{L_{1}},\frac{n_{2}}{L_{2}}\right)  $ where $n_{1}$
and $n_{2}$ are integers and $L_{1},L_{2}$ \ are the periods of the crystal
\ along the principal directions. \ Then, the field operators can be written
as
\[
\widehat{\Psi}(\mathbf{x})=\sum_{\mathbf{k}\alpha}\Psi_{\mathbf{k}\alpha
}(\mathbf{x})\text{ }\widehat{c}_{\mathbf{k}\alpha}\text{ },\text{
\ \ \  }\widehat{\Psi}^{+}(\mathbf{x})=\sum_{\mathbf{k}\alpha}%
\Psi_{\mathbf{k}\alpha}(\mathbf{x})^{+}\text{ }\widehat{c}_{\mathbf{k}\alpha
}^{+}\text{ },
\]
where $\alpha$ is the spin projection quantum number. \ Therefore, the free
propagator can be evaluated from the usual formula \cite{fetter}:%
\begin{equation}
iG_{\alpha\beta}(\mathbf{x}t,\mathbf{x^{\prime}}t^{\prime})=\frac{\langle
\Psi_{0}|T[\widehat{\Psi}_{H\alpha}(\mathbf{x}t)\widehat{\Psi}_{H\beta}%
^{+}(\mathbf{x^{\prime}}t^{\prime})]|\Psi_{0}\rangle}{\langle\Psi_{0}|\Psi
_{0}\rangle}\text{ }, \label{Definicion_Func_Green_con_Interaccion}%
\end{equation}
where $|\Psi_{0}\rangle$ is the non-interacting ground state of the model and
$\ T$ means the chronological operator.

\ The free ground state to be considered will be one in which all
the electron wave functions are filled up to some \ value of the
Fermi energy $\epsilon _{F}.$ \ Since \ the considered problem
lacks the rotational invariance, there is not a unique Fermi wave
vector $k_{F}$ associated to all the highest filled electron
states.  \ As usual, the field operators can be rewritten in terms
of the operators creating and destroying electrons above the Fermi
energy \ or holes
below it, in the following way :%
\begin{align}
\widehat{\Psi}_{S}(\mathbf{x})  &  =\sum_{\epsilon(k)>\epsilon_{F}}\Psi_{\mathbf{k}%
\lambda}(\mathbf{x})\widehat{a}_{\mathbf{k}\lambda}+\sum_{\epsilon(k)<\epsilon_{F}}%
\Psi_{\mathbf{k}\lambda}(\mathbf{x})\widehat{b}_{-\mathbf{k}\lambda}^{+}\text{
},
\label{Operador_Campo_Rep_Interac}%
\end{align}
in which the usual definitions are given of the creation of holes operators:
$\widehat{b}_{-\mathbf{k}\lambda}^{+}$ as being equal to the annihilation
operator of electrons at momenta $\mathbf{k}$ for energies lower than Fermi
one. The new annihilation operators satisfy \ $\ \widehat{a}_{\mathbf{k}\beta
}|\Phi_{0}\rangle=0$ $\ $and $\widehat{b}_{-\mathbf{k}\beta}|\Phi_{0}%
\rangle=0$ , since there are no electrons over, nor holes below
the Fermi level. Then, the expression for the valence electron
free propagator can be evaluated following the usual steps in the
standard form \cite{fetter}
\begin{widetext}
\begin{equation}
G_{\alpha\beta}^{0}(\mathbf{x}t,\mathbf{x^{\prime}}t^{\prime})=\text{$\hbar$%
}\delta_{\alpha\beta}\sum_{\mathbf{k}}\int_{-\infty}^{\infty}\frac{d\omega
}{2\pi}\frac{\Psi_{\mathbf{k}\alpha}(\mathbf{x})\Psi_{\mathbf{k}\beta}^{\ast
}(\mathbf{x^{\prime}})e^{-i\omega(t-t^{\prime})}}{\text{$\hbar$}%
\omega-\epsilon(\mathbf{k})+i\eta.sgn(\epsilon(\mathbf{k})-\epsilon_{F}%
)}\text{ }. \label{Func_Green_Prop_Fermion}%
\end{equation}
\end{widetext}


\section{Proper polarization}

\begin{figure}[h]
\begin{center}
\epsfig{figure=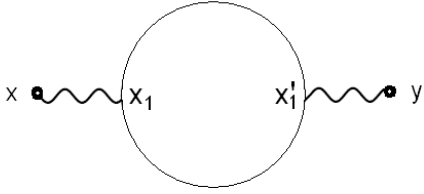,width=7cm}
\end{center}
\caption{ The first correction to the screened Coulomb potential.
The relative difficulty in the present evaluation of this loop is
produced by the lack of translational invariance of the lattice.
In the momentum representation the impulse entering in the input
line is in general different than the one in the outgoing line.
However, the reduced translational invariance of the problem will
assure the conservation of the quasi-momentum as reduced to the
first Brilouin zone. }%
\label{lazossumados}%
\end{figure}

This Section will consider the evaluation of a formula for the
polarization function. Further, these expression will be employed
to evaluate the screened Coulomb potential and the kernel of the
Bethe Salpeter bound state equation. The lowest order contribution
to the proper polarization is given by the one loop term \ formed
by two electron propagators. The associated Feynman diagram is
illustrated in the figure \ref{lazossumados} as the amputated loop
and its analytic expression can be written as follows

\begin{align}
P(x,x^{\prime})_{\lambda\lambda^{\prime}\mu\mu^{\prime}} & =G_{\lambda\mu}%
^{0}(x,x^{\prime}) \ G_{\mu^{\prime}\lambda^{\prime}}^{0}(x^{\prime},x)\text{
},\label{Polarizacion_Producto_Dos_Fun_Green}\\
& =G^{0}(x,x^{\prime}) \ G^{0}(x^{\prime},x)\delta_{\lambda\mu}\delta
_{\mu^{\prime}\lambda^{\prime}}\text{ },\nonumber\\
& =P(x,x^{\prime})\text{ }\delta_{\lambda\mu}\delta_{\mu^{\prime}%
\lambda^{\prime}}\text{ },
\end{align}
where for writing it, the formerly defined compact notation $\ x=(\mathbf{x}%
,t)$ has been employed.

  After substituting (\ref{Func_Green_Prop_Fermion}) it follows
\begin{widetext}
\begin{equation}
P(x,x^{\prime})=\text{$\hbar$}^{2}\int_{-\infty}^{\infty}\frac{d\omega}{2\pi
}\int_{-\infty}^{\infty}\frac{d\omega^{\prime}}{2\pi}\sum_{\mathbf{k}}%
\sum_{\mathbf{k^{\prime}}}\frac{e^{-i(\omega-\omega^{\prime})(t-t^{\prime}%
)}\Psi_{\mathbf{k}}(\mathbf{x})\Psi_{\mathbf{k}}^{\ast}(\mathbf{x^{\prime}%
})\Psi_{\mathbf{k^{\prime}}}(\mathbf{x^{\prime}})\Psi_{\mathbf{k^{\prime}}%
}^{\ast}(\mathbf{x})}{[\text{$\hbar$}\omega-\widetilde{\epsilon}%
(\mathbf{k})][\text{$\hbar$}\omega^{\prime}-\widetilde{\epsilon}%
(\mathbf{k^{\prime}})]}\text{ }, \label{Polarizacion_Espacio_Coordenadas}%
\end{equation}
\end{widetext} where it has been defined
\begin{equation}
\widetilde{\epsilon}(\mathbf{k})=\epsilon(\mathbf{k})-i \ \eta\ sgn(\epsilon
(\mathbf{k})-\epsilon_{F})\text{ } . \label{Repres_energ_etha+}%
\end{equation}

As it can be observed,  the polarization has explicit dependence
on two spatial arguments. This is a natural outcome since the
lattice potential breaks the full translational symmetry. \
Therefore, the model under consideration does not have a
continuous translational invariance and the technical evaluation
of the dielectric properties turn out to be more cumbersome than
for the homogeneous systems. On another hand, since the
Hamiltonian of the system is time invariant the time dependence of
the polarization will be only through the time differences.

Thus,  let us determine the  Fourier transform of
 $P(x,x^{\prime})$ over the two spatial arguments and the time. The
Fourier variables will be associated according to the rules\ \ \
$\mathbf{x}$ \ $\rightarrow$\ \ $\mathbf{q}$ $,$ \
$\mathbf{x^{\prime}}$\ \ $\rightarrow$\ \ $\mathbf{q^{\prime}}$ \
and \ $(t-t^{\prime})$\ $\rightarrow$ \ $\omega^{e}$, and the
convention to be
used for the Fourier transforms of $P(\mathbf{x},\mathbf{x^{\prime}%
},t-t^{\prime})$ \ is explicitly defined by the relation:
\begin{widetext}
\[
P(\mathbf{q},\mathbf{q^{\prime}},\omega^{e})=\int d^{2}\mathbf{x}\int
d^{2}\mathbf{x^{\prime}}\int_{-\infty}^{\infty}d(t-t^{\prime})e^{-i\mathbf{q}%
.\mathbf{x}}e^{-i\mathbf{q^{\prime}}.\mathbf{x^{\prime}}}e^{i.\omega
^{e}(t-t^{\prime})}P(\mathbf{x},\mathbf{x^{\prime}},t-t^{\prime})\text{ }.
\]
\end{widetext}
Substituting
$P(\mathbf{x},\mathbf{x^{\prime}},t-t^{\prime})$ by its expression
(\ref{Polarizacion_Espacio_Coordenadas}), conveniently reordering
the operations and employing the Dirac delta definition
$\delta(p)=\frac {1}{2\pi}\int_{-\infty}^{\infty}dx\text{
}e^{i\text{ }p\text{.}x}\text{ }, $ results in:
\begin{widetext}
\begin{align}
P(\mathbf{q},\mathbf{q^{\prime}},\omega^{e})  &  =\text{$\hbar$}^{2}%
\sum_{\mathbf{k}}\sum_{\mathbf{k^{\prime}}}\left\{  \int d^{2}\mathbf{x}\int
d^{2}\mathbf{x^{\prime}}e^{-i\mathbf{q}.\mathbf{x}-i\mathbf{q^{\prime}%
}.\mathbf{x^{\prime}}}\Psi_{\mathbf{k}}(\mathbf{x})\Psi_{\mathbf{k}}^{\ast
}(\mathbf{x^{\prime}})\Psi_{\mathbf{k^{\prime}}}(\mathbf{x^{\prime}}%
)\Psi_{\mathbf{k^{\prime}}}^{\ast}(\mathbf{x})\right\} \nonumber\\
&  \int_{-\infty}^{\infty}\frac{d\omega}{2\pi}\frac{1}{[\text{$\hbar$}%
\omega-\widetilde{\epsilon}(\mathbf{k})][\text{$\hbar$}(\omega-\omega
^{e})-\widetilde{\epsilon}(\mathbf{k^{\prime}})]}. \label{P}%
\end{align}
\end{widetext}

Let us decompose an arbitrary vector of the plane as the superposition of \ a
\ translation vector in the direct lattice \ and a vector $\mathbf{z}$
contained in the elementary Wigner-Seitz cell. \ \ Then, for any vectors
\ $\mathbf{x}$ and $\mathbf{x^{\prime}}$ is possible to write
\begin{align*}
\mathbf{x}  &  =\mathbf{z}+[\mathbf{x}]=\mathbf{z}+\mathbf{R}\text{ },\\
\mathbf{x^{\prime}}  &  =\mathbf{z^{\prime}}+[\mathbf{x^{\prime}%
}]=\mathbf{z^{\prime}}+\mathbf{R^{\prime}}\text{ }.
\end{align*}

The squared brackets $[\mathbf{x}]$ could be seen as an "integer
part" of the considered \ coordinates in the plane. That is
$[\mathbf{x}]=\mathbf{R}$ \ where $\mathbf{R}$ is a lattice vector
\ and $\mathbf{z}\in W$ in which $W$ is the elementary cell of the
crystal. \ After that, \ the above expression for
$P(\mathbf{q},\mathbf{q^{\prime}},\omega^{e})$ can be simplified
by substituting the \ integrations through all the plane over the
variables $\mathbf{x}$ or $\mathbf{x^{\prime}}$ by integrals over
the elementary cells
as:%
\begin{equation}
\int d^{2}\mathbf{x}\digamma(\mathbf{x})=\int_{W}d^{2}\mathbf{z}%
\sum_{\mathbf{R}}\digamma(\mathbf{z}+\mathbf{R})\text{ }. \label{40}%
\end{equation}
After defining the quantity
\[
I=\int d^{2}\mathbf{x}\int d^{2}\mathbf{x^{\prime}}e^{-i\mathbf{q}%
.\mathbf{x}-i\mathbf{q^{\prime}}.\mathbf{x^{\prime}}}\Psi_{\mathbf{k}%
}(\mathbf{x})\Psi_{\mathbf{k}}^{\ast}(\mathbf{x^{\prime}})\Psi
_{\mathbf{k^{\prime}}}(\mathbf{x^{\prime}})\Psi_{\mathbf{k^{\prime}}}^{\ast
}(\mathbf{x})\text{ },
\]
with the help of the Bloch functions property
\begin{align}
\Psi_{\mathbf{k}}(\mathbf{x})  &  =\Psi_{\mathbf{k}}(\mathbf{z}+\mathbf{R}%
)\text{ \ \ }=e^{i\mathbf{k}.\mathbf{R}}\text{ }\Psi_{\mathbf{k}}%
(\mathbf{z})\text{ }, \label{Autofunciones_Condicion_Bloch}%
\end{align}
and applying (\ref{40}),\ it can be written
\begin{widetext}
\begin{align*}
I  &  =\int_{W}d^{2}\mathbf{z}\int_{W}d^{2}\mathbf{z^{\prime}}\Psi
_{\mathbf{k}}(\mathbf{z})\Psi_{\mathbf{k}}^{\ast}(\mathbf{z^{\prime}}%
)\Psi_{\mathbf{k^{\prime}}}(\mathbf{z^{\prime}})\Psi_{\mathbf{k^{\prime}}%
}^{\ast}(\mathbf{z})\exp\left[  -i\mathbf{q}.\mathbf{z}-i\mathbf{q^{\prime}%
}.\mathbf{z^{\prime}}\right] \\
&  \sum_{\mathbf{R}}\exp\left[  i(\mathbf{k}-\mathbf{k^{\prime}}%
-\mathbf{q}).\mathbf{R}\right]  \sum_{\mathbf{R^{\prime}}}\exp\left[
-i(\mathbf{k}-\mathbf{k^{\prime}}+.\mathbf{q^{\prime}}).\mathbf{R^{\prime}%
}\right]  \text{ }.
\end{align*}
\end{widetext}

The following relation allow to simplify the last equation%
\begin{equation}
\sum_{\mathbf{R}}e^{i(\mathbf{k}-\mathbf{k^{\prime}}-\mathbf{q}).\mathbf{R}%
}=N\text{ }\delta^{(\mathbf{Q})}(\mathbf{k}-\mathbf{k^{\prime}}-\mathbf{q}%
)\text{ }, \label{41}%
\end{equation}
where $\delta^{(\mathbf{Q})}$ is the Kronecker delta in the space of the wave
vectors $\mathbf{k}$, defined in the first Brillouin zone. The super-index
means that it is periodically extended outside the zone. That means
$\ \ \ \delta^{(\mathbf{Q})}(\mathbf{k}-\mathbf{k^{\prime}}-\mathbf{q,0}%
)=\delta^{(\mathbf{Q})}(\mathbf{k}-\mathbf{k^{\prime}}+\mathbf{q}^{\prime
}+\mathbf{Q,0})$, with $\mathbf{Q}$ being an arbitrary vector of the
reciprocal lattice.\ \ Then, making use of (\ref{41}), the integral $I$
reduces to
\begin{widetext}
\begin{align*}
I  &  =N^{2}\delta^{(\mathbf{Q})}(\mathbf{k}-\mathbf{k^{\prime}}%
-\mathbf{q,0})\delta^{(\mathbf{Q}%
\acute{}%
)}(\mathbf{k}-\mathbf{k^{\prime}}+\mathbf{q^{\prime},0})\int_{W}%
d^{2}\mathbf{z}\int_{W}d^{2}\mathbf{z^{\prime}}\Psi_{\mathbf{k}}%
(\mathbf{z})\Psi_{\mathbf{k}}^{\ast}(\mathbf{z^{\prime}})\\
&  \Psi_{\mathbf{k^{\prime}}}(\mathbf{z^{\prime}})\Psi_{\mathbf{k^{\prime}}%
}^{\ast}(\mathbf{z})\exp\left[  -i\mathbf{q}.\mathbf{z}-i\mathbf{q^{\prime}%
}.\mathbf{z^{\prime}}\right]  \text{ }.
\end{align*}
\end{widetext}
Substituting this relation in \ the polarization expression
(\ref{P}),
\ grouping the \ terms having dependence on $\mathbf{k}$, $\mathbf{k^{\prime}%
}$ and after performing the sum over $\mathbf{k^{\prime}}$, \ the following
formula arises for the polarization loop
\begin{widetext}
\begin{gather}
P(\mathbf{q},\mathbf{q^{\prime}},\omega^{e})=\text{$\hbar$}^{2}N^{2}\frac
{A}{(2\pi)^{2}}\delta^{(\mathbf{Q})}(\mathbf{q}+\mathbf{q^{\prime},0)}%
\int_{-\infty}^{\infty}\frac{d\omega}{2\pi}\int_{W}d^{2}\mathbf{z}\int
_{W}d^{2}\mathbf{z^{\prime}}e^{-i\mathbf{q}.\mathbf{z}-i\mathbf{q^{\prime}%
}.\mathbf{z^{\prime}}}\label{Pol_intermedio1}\\
\text{ \ \ \ \ \ \ \ \ \ \ \ \ \ \ \ \ \ \ \ \ }\int d^{2}\mathbf{k}%
\Psi_{\mathbf{k}}(\mathbf{z})\Psi_{\mathbf{k}}^{\ast}(\mathbf{z^{\prime}}%
)\Psi_{\mathbf{k}-\mathbf{q}}(\mathbf{z^{\prime}})\Psi_{\mathbf{k}-\mathbf{q}%
}^{\ast}(\mathbf{z})\frac{1}{[\text{$\hbar$}\omega-\widetilde{\epsilon
}(\mathbf{k})][\text{$\hbar$}(\omega-\omega^{e})-\widetilde{\epsilon
}(\mathbf{k}-\mathbf{q})]}\text{ }.\nonumber
\end{gather}

After performing the frequency integral over $w$, it also follows
\begin{gather}
P(\mathbf{q},\mathbf{q^{\prime}},\omega^{e})=(-\frac{\text{$\hbar$}}{i}
)N^{2}\delta(\mathbf{q}+\mathbf{q^{\prime})}\int_{W}d^{2}\mathbf{z}\int
_{W}d^{2}\mathbf{z^{\prime}\exp}\left[  -i\mathbf{q}.\mathbf{z}
-i\mathbf{q^{\prime}z^{\prime}}\right]  \int
d^{2}\mathbf{k}\Psi_{\mathbf{k}
}(\mathbf{z})\Psi_{\mathbf{k}}^{\ast}(\mathbf{z^{\prime}})\nonumber\\
\Psi_{\mathbf{k}-\mathbf{q}}(\mathbf{z^{\prime}})\Psi_{\mathbf{k}-\mathbf{q}
}^{\ast}(\mathbf{z})\left\{
\frac{\theta(\epsilon_{F}-\epsilon(\mathbf{k}
-\mathbf{q}))\theta(\epsilon(\mathbf{k})-\epsilon_{F})}{[\hbar\omega
^{e}+\epsilon(\mathbf{k}-\mathbf{q})-\epsilon(\mathbf{k})+i\eta]}-\frac
{\theta(\epsilon_{F}-\epsilon(\mathbf{k}))\theta(\epsilon(\mathbf{k}
-\mathbf{q})-\epsilon_{F})}{[\hbar\omega^{e}+\epsilon(\mathbf{k}
-\mathbf{q})-\epsilon(\mathbf{k})-i\eta]}\right\}  \text{ },
\label{Polarizacion_medias4}
\end{gather}
\end{widetext} where the periodic delta function
$\delta({\bf q})=\frac{A} {2 \pi^{2}}\delta ^{Q}({\bf q})$ has
been defined. The limited lattice translational invariance of the
problem can be made explicit. Defining the the quantity $\ \
\epsilon
(\mathbf{k}-\mathbf{q})-\epsilon(\mathbf{k})=\epsilon_{\mathbf{k,q}}$
(for which the properties of the energy spectrum
(\ref{Expresion_Espectro_Energia})
$\ \ \epsilon(-\mathbf{k})=\epsilon(\mathbf{k})$ and $\epsilon(\mathbf{k}%
)=\epsilon(\mathbf{k+Q}),$ implies $\epsilon_{\mathbf{k,q+Q}}=\epsilon
_{\mathbf{k,q}}$), using the  wave property $\Psi_{\mathbf{k+Q}}(\mathbf{z}%
)=\Psi_{\mathbf{k}}(\mathbf{z})$ and substituting $\mathbf{q}%
=\mathbf{\widetilde{q}}+\mathbf{Q}$ and $\ \mathbf{q^{\prime}}%
=\mathbf{\widetilde{q}^{\prime}}+\mathbf{Q^{\prime}}$ in
(\ref{Polarizacion_medias4}), the following alternative expression can be
written
\begin{widetext}
\begin{gather}
P(\mathbf{\widetilde{q}+Q},-\mathbf{\widetilde{q}}^{\prime}-\mathbf{Q^{\prime
}},\omega^{e})=(-\frac{\text{$\hbar$}}{i})N^{2}\delta(\mathbf{\widetilde{q}%
}-\mathbf{\mathbf{\widetilde{q}}^{\prime})}\int_{W}d^{2}\mathbf{z}\int
_{W}d^{2}\mathbf{z^{\prime}\mathbf{\exp}\left[  -i\mathbf{\widetilde{q}%
}.\mathbf{z}+i\mathbf{\mathbf{\widetilde{q}}^{\prime}}.\mathbf{z^{\prime}%
}\right]  }\nonumber\\
\mathbf{\exp}\left[  -i\mathbf{Q}.\mathbf{z}+i\mathbf{Q\mathbf{^{\prime}}%
}.\mathbf{z^{\prime}}\right]  \int d^{2}\mathbf{k}\Psi_{\mathbf{k}}%
(\mathbf{z})\Psi_{\mathbf{k}}^{\ast}(\mathbf{z^{\prime}})\Psi_{\mathbf{k}%
-\mathbf{\widetilde{q}}}(\mathbf{z^{\prime}})\Psi_{\mathbf{k}%
-\mathbf{\widetilde{q}}}^{\ast}(\mathbf{z}) \nonumber \\
\left\{  \frac{\theta(\epsilon_{F}-\epsilon(\mathbf{k}-\mathbf{\widetilde{q}%
}))\theta(\epsilon(\mathbf{k})-\epsilon_{F})}{[\hbar\omega^{e}+\epsilon
_{k,\mathbf{\widetilde{q}}}+i\eta]}-\frac{\theta(\epsilon_{F}-\epsilon
(\mathbf{k}))\theta(\epsilon(\mathbf{k}-\mathbf{\widetilde{q}})-\epsilon_{F}%
)}{[\hbar\omega^{e}+\epsilon_{k,\mathbf{\widetilde{q}}}-i\eta^{\prime}%
]}\right\}  \text{ }.\label{Polarizacion_Q_Q'_Kronecker}\\
=\delta(\mathbf{\widetilde{q}}-\mathbf{\mathbf{\widetilde{q}}^{\prime}%
)}P(\mathbf{\widetilde{q}},\mathbf{Q},\mathbf{Q^{\prime}},\omega^{e})\text{ }
\label{Prop_Pol}%
\end{gather}
\end{widetext}

The lattice invariance of the problem, is reflected in this relation by the
delta function evaluated in the difference between the reduced momenta. Thus,
the limited translation invariance is able to assure the conservation of the
reduced momentum in the loops, but it can't avoid the changes over the
entering \ and outgoing reciprocal lattice vectors $Q$.

The property (\ref{Prop_Pol})\ will be of help for the simplification of the
various terms \ in what follows. For its use is helpful to eliminate the
products of Heaviside functions. Employing the property $\theta(x)=1-\theta
(-x)$ in the expression for $P(\mathbf{q},\mathbf{q^{\prime}},\omega^{e})$
defined in (\ref{Polarizacion_medias4}), the product of Heaviside functions
can be transformed in sums in arriving to the alternative formula
\begin{widetext}
\begin{gather}
P(\mathbf{q},\mathbf{q^{\prime}},\omega^{e})=\left(  -\frac{\text{$\hbar$}}%
{i}\right)  N^{2}\delta(\mathbf{q}+\mathbf{q^{\prime})}\int_{W}d^{2}%
\mathbf{z}\int_{W}d^{2}\mathbf{z^{\prime}\exp}\left[  -i\mathbf{q}%
.\mathbf{z}-i\mathbf{q^{\prime}z^{\prime}}\right]  \int d^{2}\mathbf{k}%
\Psi_{\mathbf{k}}(\mathbf{z})\Psi_{\mathbf{k}}^{\ast}(\mathbf{z^{\prime}%
})\nonumber\\
\Psi_{\mathbf{k}-\mathbf{q}}(\mathbf{z^{\prime}})\Psi_{\mathbf{k}-\mathbf{q}%
}^{\ast}(\mathbf{z})\left\{  \frac{\theta(\epsilon_{F}-\epsilon(\mathbf{k}%
-\mathbf{q}))}{[\hbar\omega^{e}+\epsilon_{k,\mathbf{\widetilde{q}}}+i\eta
]}-\frac{\theta(\epsilon_{F}-\epsilon(\mathbf{k}))}{[\hbar\omega^{e}%
+\epsilon_{k,\mathbf{\widetilde{q}}}-i\eta^{\prime}]}\right\}  \text{ }.
\end{gather}
\end{widetext} The conservation of the reduced quasi-momentum can
again be made explicit  by
writing this relation in the form:%
\[
P(\mathbf{\widetilde{q}+Q},-\mathbf{\widetilde{q}}^{\prime}-\mathbf{Q^{\prime
}},\omega^{e})=\delta(\mathbf{\widetilde{q}-\mathbf{\widetilde{q}}^{\prime}%
})P(\mathbf{\widetilde{q},Q},\mathbf{Q^{\prime}},\omega^{e})\text{ },
\]
where $\mathbf{\widetilde{q}}$ is the reduced wave vector, and it
has been defined:
\begin{widetext}
\begin{multline}
P(\mathbf{\widetilde{q},Q},\mathbf{Q^{\prime}},\omega^{e})=\left(
-\frac{\text{$\hbar$}}{i}\right)  N^{2}\int_{W}d^{2}\mathbf{z}\int_{W}%
d^{2}\mathbf{z^{\prime}\mathbf{\exp}\left[  -i\mathbf{\widetilde{q}%
}.(\mathbf{z-z^{\prime})}\right]  \exp}\left[  -i\mathbf{Q}.\mathbf{z}%
+i\mathbf{Q\mathbf{^{\prime}}}.\mathbf{z^{\prime}}\right]  \int d^{2}%
\mathbf{k}\\
\Psi_{\mathbf{k}}(\mathbf{z})\Psi_{\mathbf{k}}^{\ast}(\mathbf{z^{\prime}}%
)\Psi_{\mathbf{k}-\mathbf{\widetilde{q}}}(\mathbf{z^{\prime}})\Psi
_{\mathbf{k}-\mathbf{\widetilde{q}}}^{\ast}(\mathbf{z})\left\{  \frac
{\theta(\epsilon_{F}-\epsilon(\mathbf{k}-\mathbf{\widetilde{q}}))}%
{[\hbar\omega^{e}+\epsilon_{k,\mathbf{\widetilde{q}}}+i\eta]}-\frac
{\theta(\epsilon_{F}-\epsilon(\mathbf{k}))}{[\hbar\omega^{e}+\epsilon
_{k,\mathbf{\widetilde{q}}}-i\eta^{\prime}]}\right\}  \text{ }.
 \label{pf}
\end{multline}
\end{widetext}

\section{Screened Coulomb interaction.}
Let now consider the evaluation of the Fourier transform of the
screened Coulomb potential, associated to the partially filled
band model for the valence electrons in the $HTc$ superconductors.
\ The bare potential will be
represented in the form%
\begin{align}
U_{0}(x_{1},x_{2})  &  =V_{0}(\mathbf{x}_{1}-\mathbf{x}_{2})\delta(t_{1}%
-t_{2})\text{ },\label{Potencial_Coulomb_Delta_Tiempo}\\
V_{0}(\mathbf{x}_{1}-\mathbf{x}_{2})  &  =\frac{e^{2}}{4\pi\varepsilon
_{0}\left\vert \mathbf{x}_{1}-\mathbf{x}_{2}\right\vert },
\end{align}
where the compact notation $x\equiv(\mathbf{x,}t)$ \ continues being used in
what follows \cite{fetter}. Then, the formula:%
\[
\frac{e^{2}}{4\pi\varepsilon_{0}\left\vert \mathbf{x}-\mathbf{x^{\prime}%
}\right\vert }=\int\int\frac{d^{2}\mathbf{q}}{(2\pi)^{2}}\frac{e^{2}%
}{2\varepsilon_{0}\left\vert \mathbf{q}\right\vert }e^{i\mathbf{q}%
.(\mathbf{x}-\mathbf{x^{\prime}})}\text{ },
\]
defines the Fourier transform of $V_{0}(\mathbf{x}_{1}-\mathbf{x}_{2})$ as
\[
V_{0}(\mathbf{q})=\frac{e^{2}}{2\varepsilon_{o}\left\vert \mathbf{q}%
\right\vert }.
\]
Because the full translation invariance, the Fourier transform
over the two spatial indices and the temporal difference of the
main magnitudes being considered (polarization, screened
potential, etc.) \ will be taken. These transformations will be
written for any \ such of these quantities, let us say $Q$, in the
notation
\begin{equation}
Q(x,x^{\prime})=\int\frac{d^{2}\mathbf{q}}{(2\pi)^{2}}\frac{d^{2}%
\mathbf{q}^{\prime}}{(2\pi)^{2}}\frac{d\omega}{2\pi}e^{i\mathbf{q}.\mathbf{x}%
}e^{i\mathbf{q}^{\prime}.\mathbf{x}^{\prime}}e^{-i\omega(t_{1}-t_{2}%
)}Q(\mathbf{q},\mathbf{q}^{\prime},\omega)\text{ }. \label{FT}%
\end{equation}
Let's evaluate now the ladder approximation series for the \
interaction potential $U(x,y)_{\alpha\beta\delta\gamma}$ between
two valence electrons. \ The \ first contribution to this quantity
is represented as a Feynman diagram in the figure
\ref{lazossumados}. The result have the expression
\begin{widetext}
\begin{align}
U(x,y)_{\alpha\beta\delta\gamma}  &  =U_{0}(x,y)_{\alpha\beta\delta\gamma
}+(\frac{i}{\hbar})(-1)\int d^{3}x_{1}\int d^{3}x_{1}^{\prime}U_{0}%
(x,x_{1})_{\alpha\beta\lambda\lambda^{\prime}}P(x_{1},x_{1}^{\prime}%
)_{\lambda\lambda^{\prime}\mu\mu^{\prime}}U_{0}(x_{1}^{\prime},y)_{\mu
\mu^{\prime}\delta\gamma}+...\text{ }\label{U(x,y)=transformada_general}\\
&  =\widehat{U}_{0}+\sum_{m=1}^{\infty}(-\frac{i}{\hbar})^{m}(\widehat{U}%
_{0}\widehat{P}\text{ })^{m}\widehat{U}_{0},
\end{align}
\end{widetext}
where $U_{0}(x,y)_{\alpha\beta\delta\gamma}$ is the bare Coulomb
kernel. In this relation $\widehat{U}_{0}$ and $\widehat{P}$
indicate the functional kernels \ clearly defined by the Coulomb
potential \ and the polarization function in the first term of the
expansion. The bare potential  has the explicit form
\[
U_{0}(x,y)_{\alpha\beta\delta\gamma}=V_{0}(\mathbf{x}-\mathbf{y}%
)\delta_{\alpha\beta}\delta_{\delta\gamma}\delta(t_{x}-t_{y}).
\]
Taking the Fourier transform of \ $P$ \ \ according to \ (\ref{FT}) and
evaluating the integrals \ for a general contribution, the following
expression for the Fourier transform of the polarization function arises
\begin{widetext}
\begin{align}
U(\mathbf{q},\mathbf{q^{\prime}},\omega^{e})  &  =(2\pi)^{2}\delta^{p}
(\mathbf{q+q^{\prime}})V_{0}(\mathbf{q})+(\frac{i}{\hbar})(-1)V_{0}%
(\mathbf{q})P(\mathbf{q},\mathbf{q^{\prime}},\omega^{e})V_{0}%
(-\mathbf{q^{\prime}})+.\nonumber\\
&  ..+(\frac{i}{\hbar})^{N+1}(-1)^{N+1}\int\frac{d^{2}\mathbf{q_{1}}}%
{(2\pi)^{2}}\int\frac{d^{2}\mathbf{q_{2}}}{(2\pi)^{2}}...\int\frac
{d^{2}\mathbf{q_{N}}}{(2\pi)^{2}}V_{0}(\mathbf{q})P(\mathbf{q},\mathbf{q_{1}%
},\omega^{e})V_{0}(-\mathbf{q_{1}})\times\nonumber\\
&  P(-\mathbf{q_{1}},\mathbf{q_{2}},\omega^{e})V_{0}(-\mathbf{q_{2}%
})...P(-\mathbf{q_{N}},\mathbf{q^{\prime}},\omega^{e})V_{0}(-\mathbf{q^{\prime
}})\text{ },
\end{align}
\end{widetext} where $\delta^{p}$ defines the Dirac delta function
for the whole plane $\delta^{p}({\bf q})= \delta({\bf q}) \
\delta_{Q,-Q^{\prime}}$ and $\delta({ \bf q})$ is the already
defined periodic delta function.

\section{Tight-Binding approximation for Bethe-Salpeter two particle kernel}

Let us \ pass now to implement the tight binding approximation for the
evaluation of the screened Coulomb potential. \ Consider\ for this purpose the
Bloch theorem $\ $property $\ \Psi_{\mathbf{k}}(\mathbf{x})=\exp
(i\mathbf{kx})$ $u_{\mathbf{k}}(\mathbf{x})$, \ where the $u_{\mathbf{k}%
}(\mathbf{x})$ \ are periodic functions in the lattice. \ Applying
this relation to the wave functions (\ref{Func_Bloch_planteam})
and substituting
the explicit form of the Gaussian orbitals (\ref{orb_gaus}), it follows%
\begin{widetext}
\begin{eqnarray}
\Psi_{\mathbf{k}}(\mathbf{x})&=&\frac{1}{\sqrt{N_{\mathbf{k}}}}\sum_{\mathbf{R}%
}e^{i\mathbf{k}.\mathbf{R}}\varphi(\mathbf{x}-\mathbf{R}),  \nonumber \\
&=&\exp(i\mathbf{kx}%
)\frac{1}{\sqrt{N_{\mathbf{k}}}}\sum_{\mathbf{R}}\exp\left(  i\mathbf{k}%
.(\mathbf{R-x)}\right)  \frac{1}{\sqrt{N_{at}}}\exp\left(
-\frac{\left( \mathbf{x-R}\right)  ^{2}}{2a^{2}}\right)  \text{ }.
\end{eqnarray}
\end{widetext}
 \ Now, \ let us assume that the Gaussian orbitals \
have \ very small overlapping integrals for nearest neighbor
sites, \ allowing to consider $a<<p$. \ \ Then, it can be written
\[
u_{\mathbf{k}}(\mathbf{x})=\frac{1}{\sqrt{N_{\mathbf{k}}
N_{at}}}\sum_{\mathbf{R}}\exp\left(
i\mathbf{k}.(\mathbf{R-x)}\right) \exp\left( -\frac{\left(
\mathbf{x-R}\right) ^{2}}{2a^{2}}\right).
\]
But, the \ values of the functions $u_{\mathbf{k}}(\mathbf{x})$ will be
significant only \ for $\left\vert \mathbf{x-R}\right\vert \approx a$ \ and
thus the exponential $\exp\left(  i\mathbf{k}.(\mathbf{R-x)}\right)  \approx1
$. \ Therefore, \ the functions $u_{\mathbf{k}}(\mathbf{x})$ \ almost will not
depend on the wave vectors $\mathbf{k}$. \ This can be observed \ by
substituting the first exponential \ by the unit in $u_{\mathbf{k}}%
(\mathbf{x})$. \ Assuming this approximation, (\ref{Nk=N}) \ writes according
to
\begin{equation}
u_{\mathbf{k}}(\mathbf{x})\approx u(\mathbf{x})=\frac{1}{\sqrt{NN_{at}}}%
\sum_{\mathbf{R}}\exp\left(  -\frac{\left(  \mathbf{x-R}\right)  ^{2}}{2a^{2}%
}\right)  \text{ }. \label{u(k)}%
\end{equation}

Substituting now in the polarization expression (\ref{pf}) the relation
$u_{\mathbf{k}}(\mathbf{x})\approx u(\mathbf{x})$, gives
\begin{widetext}
\begin{align}
P(\mathbf{\widetilde{q},Q},\mathbf{Q^{\prime}},\omega^{e})  &  =\left(
-\frac{\text{$\hbar$}}{i}\right)  N^{2}\int_{W}d^{2}\mathbf{z}\text{
}u(\mathbf{z})^{2}\exp\left[  -i\mathbf{Q.z}\right]  \int_{W}d^{2}%
\mathbf{z^{\prime}u(\mathbf{z^{\prime}})^{2}\exp}\left[  \mathbf{iQ}^{\prime
}.\mathbf{.z^{\prime}}\right] \label{P(q,Q,Q,w)1}\\
&  \int d^{2}\mathbf{k}\text{ }\left\{  \frac{\theta(\epsilon_{F}%
-\epsilon(\mathbf{k}-\mathbf{\widetilde{q}}))}{[\hbar\omega^{e}+\epsilon
_{k,\mathbf{\widetilde{q}}}+i\eta]}-\frac{\theta(\epsilon_{F}-\epsilon
(\mathbf{k}))}{[\hbar\omega^{e}+\epsilon_{k,\mathbf{\widetilde{q}}}%
-i\eta^{\prime}]}\right\}  \text{ }.\nonumber
\end{align}
\end{widetext}

It can be noticed how all the exponentials in $\mathbf{k}$ and
$\mathbf{\widetilde{q}}$ were cancelled. \ We consider this outcome as the
main technical advance in this work. It allowed to obtain a close expression
for the interaction kernel and greatly simplified the evaluations. \ Now, let
us consider the quantity
\begin{equation}
\rho(\mathbf{Q})=N\int_{W}d^{2}\mathbf{z}\text{ }u(\mathbf{z})^{2}\exp\left[
-i\mathbf{Q.z}\right]  \text{ }, \label{rho}%
\end{equation}
which basically is the coefficient of the Fourier series expansion of the
periodic function $(u(\mathbf{z}))^{2}.$ For its calculation let us substitute
(\ref{u(k)}) in (\ref{rho}) to obtain
\begin{widetext}
\[
\rho(\mathbf{Q})=N\int_{W}d^{2}\mathbf{z}\text{ }\frac{1}{NN_{at}}%
\sum_{\mathbf{R}}\sum_{\mathbf{R}^{\prime}}\exp\left(  -\frac{\left(
\mathbf{z-R}\right)  ^{2}}{2a^{2}}\right)  \exp\left(  -\frac{\left(
\mathbf{z-R}^{\prime}\right)  ^{2}}{2a^{2}}\right)  \exp\left[  -i\mathbf{Q.z}%
\right]  \text{ }.
\]
\end{widetext}

Now, the double sum over the $\mathbf{R}$ \ is converted in \ a
single one since the overlapping is considered as being a very
small quantity. Thus, the overlapping integrals of orbitals being
centered in different sites can be disregarded. \ Realizing an
estimate of how small should be $a$ with respect to $p$ \ for the
overlapping to be considered small,\ it follows that when
$\frac{a}{p}=0.33$ then the overlapping integral is of the order
$\ 10$ $\%$ \ and when $\frac{a}{p}=0.289$ the integral is $5$
$\%$ . In the case of $a=0.5$ \AA \ for which $\frac{a}{p}=0.13$
the integral take the value $4.6\times10^{-7}$. In this way, up to
values near below $a=1$ \AA \ \ the approximation employed seems
to be  a reasonable one. \ \ The $\rho(\mathbf{Q})$ can be
explicitly \ evaluated as
\begin{align*}
\rho(\mathbf{Q})  &  =\frac{1}{N_{at}}\int_{W}d^{2}\mathbf{z}\sum_{\mathbf{R}%
}\exp\left(  -\frac{\left(  \mathbf{z-R}\right)  ^{2}}{a^{2}}\right)
\exp\left[  -i\mathbf{Q.z}\right] ,\\
&  =\rho(\mathbf{Q})=\frac{1}{N_{at}}\int_{W}d^{2}\mathbf{x}\text{ }%
\exp\left(  -\frac{\mathbf{x}^{2}}{a^{2}}\right)  \exp\left[  i\mathbf{Q.x}%
\right], \\
&  =\exp\left[  -\frac{a^{2}}{4}(Q_{1}^{2}+Q_{2}^{2})\right]  ,
\end{align*}
in arriving to which the change of variables
$\mathbf{z}^{\prime}=-\mathbf{z}$ was made by renaming newly
$\mathbf{z}^{\prime}$ as $\mathbf{z}$, and the property (\ref{40})
and $\exp(i\mathbf{Q.R})=1$ have been used.

\bigskip

The function $\rho(\mathbf{Q})$ is real and after substituting it in
(\ref{P(q,Q,Q,w)1}) the polarization function can be written as follows
\begin{widetext}
\begin{align}
P(\mathbf{\widetilde{q},Q},\mathbf{Q^{\prime}},\omega^{e}) & =\left(
-\frac{\text{$\hbar$}}{i}\right)  \rho^{\ast}(\mathbf{Q})\rho(\mathbf{Q}%
^{\prime})\int d^{2}\mathbf{k}\text{ }\left\{  \frac{\theta(\epsilon
_{F}-\epsilon(\mathbf{k}-\mathbf{\widetilde{q}}))}{[\hbar\omega^{e}%
+\epsilon_{k,\mathbf{\widetilde{q}}}+i\eta]}-\frac{\theta(\epsilon
_{F}-\epsilon(\mathbf{k}))}{[\hbar\omega^{e}+\epsilon_{k,\mathbf{\widetilde
{q}}}-i\eta^{\prime}]}\right\}  \text{ },\nonumber\\
& =\left(  2\pi\right)  ^{2}\left(  -\frac{\text{$\hbar$}}{i}\right)
\rho^{\ast}(\mathbf{Q})\Pi(\mathbf{\widetilde{q},}\omega^{e})\rho
(\mathbf{Q}^{\prime})\text{ },\label{P2pi}%
\end{align}
\end{widetext} where the quantity
\[
\Pi(\mathbf{\widetilde{q}})=\int\frac{d^{2}\mathbf{k}}{\left(  2\pi\right)
^{2}}\left\{  \frac{\theta(\epsilon_{F}-\epsilon(\mathbf{k}-\mathbf{\widetilde
{q}}))}{[\hbar\omega^{e}+\epsilon_{k,\mathbf{\widetilde{q}}}+i\eta]}%
-\frac{\theta(\epsilon_{F}-\epsilon(\mathbf{k}))}{[\hbar\omega^{e}%
+\epsilon_{k,\mathbf{\widetilde{q}}}-i\eta^{\prime}]}\right\}  \text{ },
\]
has been defined.

Now, let's determine the simplifications induced in the screened potential by
the tight binding approximation being imposed. For this purpose, let us
perform the change of variables $\mathbf{q}^{\prime}\mathbf{_{i}=-q_{i}}$ and
also employ \ (\ref{40}), but now applied to the momentum \ variables as%
\[
\int d^{2}\mathbf{q}F(\mathbf{q})=\int d^{2}\mathbf{\widetilde{q}}%
\sum_{\mathbf{Q}}F(\mathbf{\widetilde{q}}+\mathbf{Q}),
\]
where $\mathbf{q}=\mathbf{\widetilde{q}}+\mathbf{Q}$,\ $\mathbf{\widetilde{q}%
}$ \ is a reduced wave vectors and $\mathbf{Q}$ is an arbitrary
reciprocal lattice vector. \ After considering that any of the
polarization functions appearing has a Dirac delta in the
difference between the entering and outgoing reduced momenta it
follows
\begin{widetext}
\begin{multline}
U(\mathbf{q},\mathbf{q^{\prime}},\omega^{e})=(2\pi)^{2}\delta
(\mathbf{q+q^{\prime}})V_{0}(\mathbf{q})+(2\pi)^{2}\delta(\widetilde
{\mathbf{q}}\mathbf{+}\widetilde{\mathbf{q}}\mathbf{^{\prime}})V_{0}%
(\widetilde{\mathbf{q}}\mathbf{+Q})\rho^{\ast}(\mathbf{Q})\Pi
(\mathbf{\widetilde{q},}\omega^{e})\rho(\mathbf{Q}^{\prime})V_{0}%
(\widetilde{\mathbf{q}}\mathbf{-Q}^{\prime})+ \nonumber \\
+\sum_{N=1}.(2\pi)^{2}\delta(\widetilde{\mathbf{q}}\mathbf{+}\widetilde
{\mathbf{q}}\mathbf{^{\prime}})V_{0}(\widetilde{\mathbf{q}}\mathbf{+Q}%
)\rho^{\ast}(\mathbf{Q})\Pi(\mathbf{\widetilde{q},}\omega^{e})\left[
{\displaystyle\sum_{\mathbf{Q}_{1}}}
\rho^{\ast}(\mathbf{Q}_{1})V_{0}(\widetilde{\mathbf{q}}\mathbf{+Q}_{1}%
)\rho(\mathbf{Q}_{1})\right]  \Pi(\mathbf{\widetilde{q},}\omega^{e}%
)\times\nonumber\\
\left[
{\displaystyle\sum_{\mathbf{Q}_{2}}}
\rho^{\ast}(\mathbf{Q}_{2})V_{0}(\widetilde{\mathbf{q}}\mathbf{+Q}_{2}%
)\rho(\mathbf{Q}_{2})\right]  ...\Pi(\mathbf{\widetilde{q},}\omega^{e})\left[
%
{\displaystyle\sum_{\mathbf{Q_{N}}}}
\rho^{\ast}(\mathbf{Q_{N}})V_{0}(\widetilde{\mathbf{q}}\mathbf{+Q_{N}}%
)\rho(\mathbf{Q_{N}})\right]  \Pi(\mathbf{\widetilde{q},}\omega^{e}%
)\rho(\mathbf{Q}^{\prime})V_{0}(\widetilde{\mathbf{q}}\mathbf{-Q}^{\prime
})\text{,} \nonumber \\
=(2\pi)^{2}\delta(\mathbf{q+q^{\prime}})V_{0}(\mathbf{q})+(2\pi)^{2}%
\delta(\widetilde{\mathbf{q}}\mathbf{+}\widetilde{\mathbf{q}}\mathbf{^{\prime
}})V_{0}(\widetilde{\mathbf{q}}\mathbf{+Q})\rho^{\ast}(\mathbf{Q}%
)\Pi(\mathbf{\widetilde{q},}\omega^{e})\rho(\mathbf{Q}^{\prime})V_{0}%
(\widetilde{\mathbf{q}}\mathbf{-Q}^{\prime})+\nonumber\\
+\sum_{N=1}(2\pi)^{2}\delta(\widetilde{\mathbf{q}}\mathbf{+}\widetilde
{\mathbf{q}}\mathbf{^{\prime}})V_{0}(\widetilde{\mathbf{q}}\mathbf{+Q}%
)\rho^{\ast}(\mathbf{Q})%
{\displaystyle\sum_{\mathbf{N=1}}^{\infty}}
\left[  \Pi(\mathbf{\widetilde{q},}\omega^{e})\widetilde{V}(\widetilde
{\mathbf{q}})\right]  ^{\mathbf{N}}\Pi(\mathbf{\widetilde{q},}\omega^{e}%
)\rho(\mathbf{Q}^{\prime})V_{0}(\widetilde{\mathbf{q}}\mathbf{-Q}^{\prime
})\text{ },
\end{multline}
\end{widetext}
where \ $\widetilde{\mathbf{q}}$ and $\widetilde{\mathbf{q}^{\prime}%
}$ are the reduced quasi-momenta associated to \ $\mathbf{q}$ \
and $\mathbf{q}^{\prime}$ and  the sums over the $\mathbf{Q_{i}}$
were accommodated in order to absorb all of them \ in the quantity
that in what follows will be identified as the bare Coulomb kernel
of the Bethe Salpeter bound state equation
\begin{equation}
\widetilde{V}(\widetilde{\mathbf{q}})=%
{\displaystyle\sum_{\mathbf{Q}}}
\rho^{\ast}(\mathbf{Q})V_{0}(\widetilde{\mathbf{q}}\mathbf{+Q})\rho
(\mathbf{Q}).\label{V(q)conlas_rho}%
\end{equation}
\ It is possible to incorporate  in the second term in the last
line of the last expression for $U$,  in the appearing summation
to write%
\begin{widetext}
\begin{align}
U(\mathbf{\widetilde{\mathbf{q}}+Q},\widetilde{\mathbf{q}}\mathbf{^{\prime}%
+Q}^{\prime},\omega^{e}) &  =(2\pi)^{2}\delta(\mathbf{q+q^{\prime}}%
)V_{0}(\mathbf{q})+(2\pi)^{2}\delta(\widetilde{\mathbf{q}}\mathbf{+}%
\widetilde{\mathbf{q}}\mathbf{^{\prime}})V_{0}(\widetilde{\mathbf{q}%
}\mathbf{+Q})\rho^{\ast}(\mathbf{Q})\nonumber\\
&
{\displaystyle\sum_{\mathbf{N=0}}^{\infty}}
\left[  \Pi(\mathbf{\widetilde{q},}\omega^{e})\widetilde{V}(\widetilde
{\mathbf{q}})\right]  ^{\mathbf{N}}\Pi(\mathbf{\widetilde{q},}\omega^{e}%
)\rho(\mathbf{Q}^{\prime})V_{0}(\widetilde{\mathbf{q}}\mathbf{-Q}^{\prime
})\text{ }.\nonumber\\
&  =(2\pi)^{2}\delta(\widetilde{\mathbf{q}}\mathbf{+}\widetilde{\mathbf{q}%
}\mathbf{^{\prime}})V_{0}(\mathbf{q})\left\{  \delta_{\mathbf{Q,-Q}^{\prime}%
}+\frac{\Pi(\mathbf{\widetilde{q},}\omega^{e})\rho^{\ast}(\mathbf{Q}%
)V_{0}(\widetilde{\mathbf{q}}\mathbf{-Q}^{\prime})\rho(\mathbf{Q}^{\prime}%
)}{1-\Pi(\mathbf{\widetilde{q},}\omega^{e})\widetilde{V}(\widetilde
{\mathbf{q}})}\right\} ,\label{u(q,Q)final}%
\end{align}
\end{widetext}
where,  also,  the \ geometrical series was formally summed over.
This series
\ is only convergent \ when $\left\vert \Pi(\mathbf{\widetilde{q}}%
)\widetilde{V}(\widetilde{\mathbf{q}})\right\vert <1$. \ As it
will be the case, this condition will  not be satisfied in the
whole Brillouin zone (See section 6). \ However, in our problem
there will exist domains of the variable $\widetilde {\mathbf{q}}$
\ over which $\left\vert \Pi(\mathbf{\widetilde{q},}\omega
^{e})\widetilde{V}(\widetilde{\mathbf{q}})\right\vert <1$ and for
them the summation relation is obeyed. Therefore, after assuming
that there is no change in the analytical expression for the
considered quantity when the momenta is varied, the validity of
the resulting formula is  taken  as the analytical extension of
the values in the convergence region. In support of the above
interpretation is  also the fact that in the linear response
theory \cite{march},  the general posing of the problem \
validates  the employed formal summation formula.
 The equation (\ref{u(q,Q)final}) \ also can be written in the form%
\begin{equation}
U(\mathbf{\widetilde{\mathbf{q}}+Q},\widetilde{\mathbf{q}}\mathbf{^{\prime}%
+Q}^{\prime},\omega^{e})=(2\pi)^{2}\delta(\widetilde{\mathbf{q}}%
\mathbf{+}\widetilde{\mathbf{q}}\mathbf{^{\prime}})V_{0}(\mathbf{q}%
)\chi(\mathbf{\widetilde{q},Q,Q}%
\acute{}%
.\omega^{e})\text{ },\label{U1/epsilon}%
\end{equation}
 just defining the quantity $\chi$.

\section{ Bethe-Salpeter  Coulomb kernel and dielectric function}

   Up to now, we \ have been considering the polarization and
effective potential kernels of a non translational invariant
system. Therefore, the dependence of two spatial or momenta
arguments makes their study more involved. However, it can be
expected  that the kernel of the two valence electrons (or holes)
bound state problem, could be simplified by the conservation of
the reduced momenta in the  effective interaction. Thus, let us
evaluate the kernel of the Bethe-Salpeter equation associated to
the Coulomb interaction. \ This will be the relevant piece in the
discussion of the effects of the screening to which this work is
devoted. \ This kernel is determined by the \ matrix element of
the \ screened Coulomb potential (\ref{u(q,Q)final}) describing
the dispersion of two valence electrons.  Its analytic expression
is given by
\begin{widetext}
\begin{equation}
\Gamma(\mathbf{k,k}^{\prime},\mathbf{p,p}^{\prime})=\int d^{3}x\int d^{3}%
y\Psi_{\mathbf{kI}}^{+}(x)\Psi_{\mathbf{pI}}^{+}(y)U(x,y)\Psi_{\mathbf{p}%
^{\prime}\mathbf{I}}(y)\Psi_{\mathbf{k}^{\prime}\mathbf{I}}(x)\text{ }.
\label{lambda}%
\end{equation}
\end{widetext}

The wave functions are taken in the interaction representation and
the already
defined convention is used: $x=(\mathbf{x,}t)$ and $y=(\mathbf{y,}t%
\acute{}%
)$. \ They have the expressions
$\Psi_{\mathbf{kI}}(x)=\Psi_{\mathbf{k}}(\mathbf{x})e^{-i\omega_{\mathbf{k}}%
t}$ .\ For this calculation the simplifying static limit
$\omega^{e}=0$ will be assumed. That is, further we will only
explore the screening in the static approximation. \

In evaluating the matrix element the following \ steps were
followed: a) The effective potential was substituted by its
Fourier transform according to (\ref{FT}).\ b) The transformation
(\ref{40}) was applied to the spatial $\mathbf{x,y}$ as well as
the momentum \ integrals $\mathbf{q,
\ q}^{\prime}$, making use of the Bloch condition $\Psi_{\mathbf{k}%
}(\mathbf{z+R})=e^{i\mathbf{k.R}}\Psi_{\mathbf{k}}(\mathbf{z})$ and the
relation $\Psi_{\mathbf{k}}(\mathbf{z})=e^{i\mathbf{k.R}}u_{\mathbf{k}%
}(\mathbf{z})$. \ c) \ \ \ The \ tight- binding \ approximation
(\ref{u(k)}); that is,  $u_{\mathbf{k}}(\mathbf{z})\approx
u(\mathbf{z})$ \ not depending of $\mathbf{k}$, was implemented. \
d) \ The relation $\
\sum_{\mathbf{R}}e^{i\mathbf{k.R}}=N\delta_{\mathbf{k,0}}$ was
used. \ e) \ The
\ standard formula for \ macroscopic crystals $\ \delta_{\mathbf{k,0}}%
=\frac{(2\pi)^{2}}{A}\delta(\mathbf{k-}0)$ where A is area of the
lattice, was employed, and finally: f) \ The following expression
for $U$ was substituted
\begin{widetext}
\begin{align}
U(\mathbf{\widetilde{\mathbf{q}}+Q},\widetilde{\mathbf{q}}\mathbf{^{\prime}%
+Q}^{\prime},0)  &  =(2\pi)^{2}\delta(\widetilde{\mathbf{q}}\mathbf{+}%
\widetilde{\mathbf{q}}\mathbf{^{\prime}})V_{0}(\mathbf{q})\chi
(\mathbf{\widetilde{q},Q,Q}%
\acute{}%
,0), \label{1/epsilon}\\
\chi(\mathbf{\widetilde{q},Q,Q}%
\acute{}%
,0)  &  =\delta_{\mathbf{Q,-Q}^{\prime}}+\frac{\Pi(\mathbf{\widetilde{q}%
,0})\rho^{\ast}(\mathbf{Q})V_{0}(\widetilde{\mathbf{q}}\mathbf{-Q}^{\prime
})\rho(\mathbf{Q}^{\prime})}{1-\Pi(\mathbf{\widetilde{q},0})\widetilde
{{\Large {\mathbf{V}}}}(\widetilde{\mathbf{q}})},
\label{1/epsilon2}
\end{align}
\end{widetext}
as considered in the static limit $\omega^{e}=0 $ $.$

After the above enumerated  transformations \ the matrix
element (\ref{lambda}) \ can be \ obtained in the following form:%
\begin{widetext}
\begin{align*}
\Gamma(\mathbf{k,k}^{\prime},\mathbf{p,p}^{\prime})  &  =\delta(\omega
_{\mathbf{k}}-\omega_{\mathbf{k}^{\prime}}+\omega_{\mathbf{p}}-\omega
_{\mathbf{p}^{\prime}})\delta(\mathbf{k-k}^{\prime}+\mathbf{p-p}^{\prime
})\frac{(2\pi)^{3}N^{2}}{A^{2}}\\
&  \sum_{\mathbf{Q}}\int d^{2}\mathbf{z}\text{ }u^{2}(\mathbf{z)}%
\exp(i\mathbf{Q}.\mathbf{z)}\sum_{\mathbf{Q}^{\prime}}\int d^{2}%
\mathbf{z}^{\prime}u^{2}(\mathbf{z}^{\prime})\exp(i\mathbf{\mathbf{Q}}%
^{\prime}.\mathbf{z}^{\prime})\\
&  \exp(-i(\mathbf{k-k}^{\prime}+\mathbf{p-p}^{\prime}).\mathbf{z}^{\prime
})V_{0}(\mathbf{k-k}^{\prime}+\mathbf{Q})\chi(\mathbf{\widetilde{q},Q,Q}%
\acute{}%
)\text{ },
\end{align*}
\end{widetext}
where $\chi(\mathbf{\widetilde{q},Q,Q}%
\acute{}%
)=\chi(\mathbf{\widetilde{q},Q,Q}%
\acute{},0).$

Transforming the Dirac delta in a Kronecker one and using the
representation (\ref{rho}),  after also considering that
$\rho(\mathbf{Q})$ is a real and even
function, the previous expression \ is reduced to%
\begin{widetext}
\begin{align*}
\Gamma(\mathbf{k,k}^{\prime},\mathbf{p,p}^{\prime})  &  =\delta(\mathbf{p-p}%
^{\prime}+\mathbf{k-k}^{\prime},0)\delta(\omega_{\mathbf{k}}-\omega
_{\mathbf{k}^{\prime}}+\omega_{\mathbf{p}}-\omega_{\mathbf{p}^{\prime}}%
)\frac{(2\pi)}{A},\\
&  \sum_{\mathbf{Q}}\sum_{\mathbf{Q}^{\prime}}\rho(\mathbf{Q})\rho
(\mathbf{Q}^{\prime})V_{0}(\mathbf{k-k}^{\prime}+\mathbf{Q})\chi
(\mathbf{\widetilde{q},Q,Q}%
\acute{}%
)\text{ },\\
&  =\delta(\mathbf{p-p}^{\prime}+\mathbf{k-k}^{\prime},0)\delta(\omega
_{\mathbf{k}}-\omega_{\mathbf{k}^{\prime}}+\omega_{\mathbf{p}}-\omega
_{\mathbf{p}^{\prime}})\frac{(2\pi)}{A}\Gamma_{0}(\mathbf{\widetilde{q}%
,k,k}^{\prime}),
\end{align*}
\end{widetext}
where the quantity
$\Gamma_{0}(\mathbf{\widetilde{q},k,k}^{\prime})$ is defined.

\ Let us analyze the first term appearing in the definition of \
$\Gamma _{0}(\mathbf{\widetilde{q},k,k}^{\prime})$ containing the
Kronecker delta $\delta_{\mathbf{Q,-Q}^{\prime}}$, and which
corresponds to the Coulomb potential in the vacuum. For this
contribution, it follows
\begin{align*}
\Gamma_{0}^{(1)}(\mathbf{\widetilde{q},k,k}^{\prime})  &  =\sum_{\mathbf{Q}%
}\rho(\mathbf{Q})V_{0}(\mathbf{k-k}^{\prime}+\mathbf{Q})\sum_{\mathbf{Q}%
^{\prime}}\rho(\mathbf{Q}^{\prime})\delta_{\mathbf{Q,-Q}^{\prime}}\text{ },\\
&  =\sum_{\mathbf{Q}}\rho(\mathbf{Q})V_{0}(\mathbf{k-k}^{\prime}%
+\mathbf{Q})\rho(\mathbf{Q})\text{ },\\
&  =\widetilde{V}(\mathbf{k-k}^{\prime})=\widetilde{V}(\mathbf{\widetilde{q}%
})\text{ }.
\end{align*}

\begin{figure}[htb]
\begin{center}
\epsfig{figure=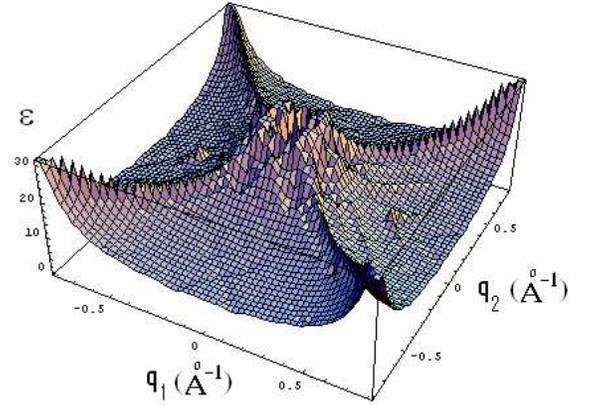,width=9cm}
\end{center}
\caption{ The dependence of the dielectric function on the quasi-momentum
$\mathbf{\widetilde{q}}$ in the Brillouin zone, for a value $a=0.95$ for the
overlapping parameter.}%
\label{grafconstdiel}%
\end{figure}

Performing a similar transformation for the second term appearing
in (\ref{1/epsilon2}), which represents the modification of the \
Coulomb
potential \ created by the \ 2D valence electrons, it follows:%
\begin{widetext}
\begin{align*}
\Gamma_{0}^{(2)}(\mathbf{\widetilde{q},k,k}^{\prime})  &  =\sum_{\mathbf{Q}%
}\sum_{\mathbf{Q}^{\prime}}\rho(\mathbf{Q})\rho(\mathbf{Q}^{\prime}%
)V_{0}(\mathbf{k-k}^{\prime}+\mathbf{Q})\frac{\Pi(\mathbf{\widetilde{q}}%
)\rho^{\ast}(\mathbf{Q})V_{0}(\widetilde{\mathbf{q}}\mathbf{-Q}^{\prime}%
)\rho(\mathbf{Q}^{\prime})}{1-\Pi(\mathbf{\widetilde{q}})\widetilde
{V}(\widetilde{\mathbf{q}})}\text{ },\\
&  =\left(  \widetilde{V}(\widetilde{\mathbf{q}})\right)  ^{2}\frac
{\Pi(\mathbf{\widetilde{q}})}{1-\Pi(\mathbf{\widetilde{q}})\widetilde
{V}(\widetilde{\mathbf{q}})}\text{ },\\
&  =\frac{\left(  \Pi(\mathbf{\widetilde{q}})\widetilde{V}(\widetilde
{\mathbf{q}})\right)  ^{2}}{\Pi(\mathbf{\widetilde{q}})}\frac{1}%
{1-\Pi(\mathbf{\widetilde{q}})\widetilde{V}(\widetilde{\mathbf{q}})}\text{ }.
\end{align*}
\end{widetext}
\ $\ \ $Now, after summing the two contributions, the following
simple result arises for the Coulomb part of the kernel
\begin{align*}
\Gamma_{0}(\mathbf{\widetilde{q},k,k}^{\prime})  &  =\Gamma_{0}^{(1)}%
(\mathbf{\widetilde{q},k,k}^{\prime})+\Gamma_{0}^{(2)}(\mathbf{\widetilde
{q},k,k}^{\prime})\text{ },\\
&  =\widetilde{V}(\mathbf{\widetilde{q}})+\frac{\left(  \Pi(\mathbf{\widetilde
{q}})\widetilde{V}(\widetilde{\mathbf{q}})\right)  ^{2}}{\Pi
(\mathbf{\widetilde{q}})}\frac{1}{1-\Pi(\mathbf{\widetilde{q}})\widetilde
{V}(\widetilde{\mathbf{q}})}\text{ },\\
&  =\frac{\widetilde{V}(\mathbf{\widetilde{q}})}{1-\Pi(\mathbf{\widetilde{q}%
})\widetilde{V}(\widetilde{\mathbf{q}})}\text{ }.
\end{align*}
Henceforth, the dielectric function related with the screening of
the bare potential has  the formula
\begin{equation}
\varepsilon(\mathbf{\widetilde{q}})=1-\Pi(\mathbf{\widetilde{q}})\widetilde
{V}(\widetilde{\mathbf{q}})\text{ }. \label{fin}%
\end{equation}
The \ evaluation of $\varepsilon(\mathbf{\widetilde{q}})$ \ was
done for the overlapping parameter having  value
\[
a=0.95\text{ \AA },
\]
and for a Fermi energy $ \epsilon_F=-0.005 \ eV$, being very
close, but below the  mid of the band. Thus, there is a very low
density of holes in the system.
 \ \ The result of the calculation plotted over the Brillouin
cell is shown in the figure \ref{grafconstdiel}.
As it can be observed the dielectric function shows high values
for special zones in the Brillouin cell that run from more than 30
near
the zone corners (like the point $\mathbf{\widetilde{q}=(}\frac{\pi}{p}%
,\frac{\pi}{p}\mathbf{))}$ down to less than 10 in other regions.
\ Note that the \ zones near the diagonals correspond to relative
high values of the dielectric response. \ The fourth order
symmetry of the result is also evidenced, although some numerical
errors due to the presence of singularities in the evaluated
integrals weakly break it. \ \ The peaks along the diagonals are
only an artifact resulting from the evaluation for a finite number
of points in the Brillouin cell. \ The obtained results momentum
dependence of the dielectric function becomes a consequence of the
2D nature of the problem. Moreover, if the half filling condition
is approached closely the dielectric constant increases
drastically (up to values of the order of 2500 were numerically
evaluated by us). Therefore, it is clear that the  Van-Hove
singularities play a central role in the obtained effect.
  An additional and important outcome  is that upon varying
the Fermi energy away from half filling ($\epsilon_F=0$), the
values of the dielectric function along the diagonals rapidly
decreases and the dependence becomes more isotropic.  That is, the
creation of holes reduces the screening of the Coulomb potential.
This property could bring a natural explanation for the firstly
rising and after decreasing of the critical temperatures in the
$HTc$ materials. The picture could be as follows: first, when the
number of hole density grows in the low density region, the strong
screening state is established and the superconductivity becomes
stronger (Tc growing). However, as the hole density is increased
even more the screening effect becomes weaker, and   cancelling
the enforcing effect of the growing density by increasing the
Coulomb repulsion. The rapid weakening effect given by the
calculations done here support this last property.

\ It should remarked that in this work we had not taken into
account the dielectric constant of the medium $\varepsilon_{b}$
(due to the polarization of the ions and atomic cores). \ It seems
that at distances of few unit cells this dielectric response could
not be fully developed, but certainly its effects could be not so
weak. Therefore, it seems reasonable to introduce a multiplicative
factor $\lambda$ ranging between 0 \ and 1, \ describing the
effective proportion of the dielectric response of the ionic and
core medium, which could be acting at small distances of the order
of the Cooper pair size. \ For large distances,  the parameter
should be equal to one and for very short ones should tend to
become smaller than one. In general, it could expected that this
effect can contribute even more to the screening of the Coulomb
interaction obtained here.
\begin{figure}[htb]
\begin{center}
\epsfig{figure=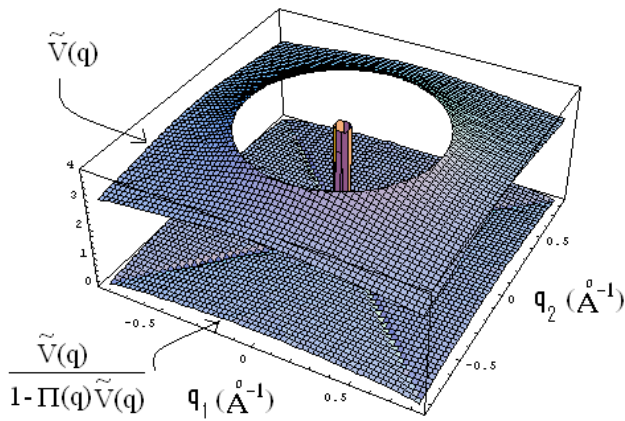,width=9cm}
\end{center}
\caption{ Plot of the screened and bare Coulomb contributions to the kernel of
the Bethe-Salpeter equation, as functions of the wave vector
$\mathbf{\widetilde{q}}$. Note the presence of an intense screening of the
Coulomb repulsion in the considered here approximation. }%
\label{grafkernel}%
\end{figure}\ \
Further, in the figure \ \ref{grafkernel}, the results of the
evaluation of the screened Coulomb kernel of the bound state
equation for two electrons (or two holes)  are plotted, in common
with the values of the bare kernel. Again the overlapping
parameter and the Fermi energy are   $\ a= 0.95$ \AA, \ $
\epsilon_F=-0.005 \ eV $. \ \ \ As it can be observed,  in the
considered approximation,  the screening effect is noticeable. \ \
Considering  that the bare Coulomb interaction produces a
repulsive potential of few $eV$ at distances of the order of the
lattice constant, it follows that the screening effect could
reduce these values down to ones of the order 0.1 $eV$ . \ But, at
such small repulsion forces other mechanisms, such as
super-exchange or strong phonon interactions \ could then create
the necessary binding forces for the pairs to form.

Finally it should be remarked that in order to approach the
present discussion to the real situation in the superconductor
materials, it seems necessary to derive a similar picture but in a
self-consistent approach. \ A concrete way for this program could
proceed in the following steps: a) To formulate the general
Coulomb interacting problem of the valence electrons but retaining
the Coulomb interaction exactly. b) Attempt to derive the here
discussed tight binding model as a kind of HF approximation of the
exact theory. c) After that, to derive and solve  the
Bethe-Salpeter equation for bounded pairs, \ which should be
expected to end including the anti-ferromagnetic interactions of
the $tj$-model as a possible bounding mechanism \cite{Zhang_Rice}.
The consideration of this program is in progress.

\begin{acknowledgments}
The support of our colleagues and friends in the Group of
Theoretical Physics of ICIMAF and the Faculty of Physics at Havana
University, where the main content of this work has been done is
deeply acknowledged. In addition, the invitation and kind
hospitality of the Condensed Matter Section of the Abdus Salam
International Centre for Theoretical Physics (ASICTP) and its Head
Prof. V. Kravtsov, allowing for a visit to the Center of one of
the authors (A.C.M.), is greatly appreciated.
\end{acknowledgments}

\end{document}